\documentstyle[aas2pp4,twoside]{article}
\input epsf
\def\s2n{S^{\prime}/N}

\def\aq{{\langle Q \rangle}}
\def\jco{{J=1-0 $^{13}$CO}}

\begin{document}
\title{The Turbulent Shock Origin of Proto--Stellar Cores}

\author{Paolo Padoan\footnote{ppadoan@cfa.harvard.edu},
\affil{Harvard-Smithsonian Center for Astrophysics, Cambridge, MA 02138}
Mika Juvela\footnote{juvela@astro.helsinki.fi},
\affil{Helsinki University Observatory,T\"ahtitorninm\"aki, P.O.Box 14,
       SF-00014 University of Helsinki, Finland}
Alyssa A. Goodman\footnote{agoodman@cfa.harvard.edu}
\affil{Harvard-Smithsonian Center for Astrophysics, Cambridge, MA 02138}
and \AA ke Nordlund\footnote{aake@astro.ku.dk}}
\affil{Astronomical Observatory and Theoretical Astrophysics Center, \\
       Juliane Maries Vej 30, DK-2100 Copenhagen, Denmark}

\begin{abstract}

The fragmentation of molecular clouds (MC) into proto--stellar cores 
is a central aspect of the process of star formation. Because of the 
turbulent nature of super--sonic motions in MCs, it has been suggested 
that dense structures such as filaments and clumps are formed by 
shocks in a turbulent flow. In this work we present strong
evidence in favor of the turbulent origin of the fragmentation
of MCs.

The most generic result of turbulent fragmentation is that dense
post shock gas traces a gas component with a smaller velocity dispersion
than lower density gas, since shocks correspond to regions of converging
flows, where the kinetic energy of the turbulent motion is dissipated.

Using synthetic maps of spectra of molecular transitions, computed
from the results of numerical simulations of super--sonic turbulence,
we show that the dependence of velocity dispersion on gas density
generates an observable relation between the rms velocity centroid 
and the integrated intensity (column density), $\sigma(V_0)-I$, which is 
indeed found in the observational data. The comparison 
between the theoretical model (maps of synthetic $^{13}$CO spectra), 
with $^{13}$CO maps from the Perseus, Rosette and Taurus MC complexes, 
shows excellent agreement in the $\sigma(V_0)-I$ relation.

The $\sigma(V_0)-I$ relation of different observational
maps with the same total rms velocity are remarkably
similar, which is a strong indication of their origin from a very
general property of the fluid equations, such as the turbulent
fragmentation process.

\end{abstract}


\section{Introduction}

The importance of turbulence in the process of star formation was recognized long 
ago (Von Weizs{\"a}cker 1951), and was discussed in a seminal paper
\nocite{von51} 
by Larson (1981). Several successive works have tried to use the 
observational data to relate different properties of MCs with the
physics of laboratory and numerical turbulence, such as power
spectra of kinetic energy \footnote{See Leung, Kutner \& Mead 1982; Myers 1983; 
Quiroga 1983;  Sanders, Scoville \& Solomon 1985;   Goldsmith \& Arquilla 1985; 
Dame et al.\ 1986; Falgarone \& P\'{e}rault 1987.}
\nocite{Leung+82,Myers83,Quiroga83}
\nocite{Sanders+85} 
\nocite{Goldsmith+Arquilla85,Dame+86,Falgarone+Perault87}, 
probability distributions of velocity and velocity differences \footnote{ 
See Scalo 1984; Kleiner \& Dickman 1985, 1987; Hobson 1992; Miesch \& 
Bally 1994; Miesch \& Scalo 1995; Lis et al. 1996; Miesch, Scalo \& 
Bally 1999.}, intermittency (Falgarone \& Phillips 1990; Falgarone
\& Puget 1995; Falgarone, Pineau Des Forets, \& Roueff 1995),
\nocite{Falgarone+Phillips90} \nocite{Falgarone+Puget95} \nocite{Falgarone+95}
and self--similarity \footnote{See Beech 1987; Bazell \& D\'{e}sert 1988, Scalo 1990; 
Dickman, Horvath \& Margulis 1990; Falgarone, Phillips \& Walker 1991; 
Zimmermann, Stutzki \& Winnewisser 1992; Henriksen 1991; 
Hetem \& Lepine 1993; Vogelaar \& Wakker 1994; Elmegreen \& Falgarone 1996.}. 
\nocite{Beech87}  \nocite{Bazell+Desert88}  \nocite{Scalo90}
\nocite{Dickman+90}  \nocite{Falgarone+91} \nocite{Zimmermann+92} 
\nocite{Henriksen91} \nocite{Hetem+Lepine93} \nocite{Vogelaar+Wakker94} 
\nocite{Elmegreen+Falgarone96}

During the last decade, numerical simulations of transonic turbulence
(Passot \& Pouquet 1987; Passot, Pouquet \& Woodward 1988; L\'{e}orat, 
Passot \& Pouquet 1990; Lee, Lele \& Moin 1991 \nocite{Lee+91}; Porter, 
Pouquet \& Woodward 1992 \nocite{Porter+94}; Kimura \& Tosa 1993; 
Porter, Woodward \& Pouquet 1994; Vazquez--Semadeni 1994; Passot, 
Vazquez--Semadeni \& Pouquet 1995) 
and highly super--sonic turbulence (Passot, V\'{a}zquez--Semadeni, Pouquet
1995; V\'{a}zquez--Semadeni, Passot \& 
Pouquet 1996; Padoan \& Nordlund 1997, 1999; Stone, Ostriker \& 
Gammie 1998; MacLow et al. 1998; Ostriker, Gammie \& Stone 1999; Padoan, Zweibel \& 
Nordlund 2000; Klessen, Heitsch \& MacLow 2000), on relatively high--resolution 
numerical grids, have become available, and very detailed comparisons 
between observational data with turbulence models of MCs have been
performed (Padoan, Jones \& Nordlund 1997; Padoan et al. 1998; Padoan \& 
Nordlund 1999; Padoan et al. 1999; Rosolowsky et al. 1999; Padoan, 
Rosolowsky \& Goodman 2000). 

Recent numerical studies of super--sonic
magneto--hydrodynamic (MHD) turbulence have brought new
understanding of the physics of turbulence. The most important
results are:

\nocite{MacLow_Puebla98} \nocite{Stone+98} \nocite{Kimura+Tosa93} 
\nocite{Lee+91} \nocite{Vazquez-Semadeni+96} \nocite{Klessen+99}
\nocite{Padoan+97ext} \nocite{Padoan+98per} \nocite{Padoan+Nordlund98MHD}
\nocite{Padoan+98cat} \nocite{Padoan+Nordlund97MHD} \nocite{Ostriker+99}
\nocite{Passot+95} \nocite{MacLow+98} \nocite{MacLow99}
\nocite{Rosolowsky+99} \nocite{Padoan+2000SCF}

\begin{itemize}
\item Super--sonic turbulence decays in approximately one dynamical 
time, independent of the magnetic field strength (Padoan \& Nordlund 1997, 1999;
MacLow et al. 1998; Stone, Ostriker \& Gammie 1998; MacLow 1999).
\item The probability distribution of gas density in isothermal turbulence is well 
approximated by a Log--Normal distribution, whose standard deviation 
is a function of the rms Mach number of the flow (Vazquez--Semadeni 1994;
Padoan 1995; Padoan, Jones \& Nordlund 1997; Scalo et al. 1998; Passot
\& Vazquez--Semadeni 1998; Nordlund \& Padoan 1999; Ostriker, Gammie \& Stone 1999).
\item Super--sonic isothermal turbulence generates a complex system 
of shocks which fragment the gas very efficiently into high density
sheets, filaments, and cores (this is the general result of any numerical 
simulation of super--sonic turbulence). 
\item Super--Alfv\'{e}nic turbulence provides a good description of the
dynamics of MCs, and an explanation for the origin of dense cores with
magnetic field strength consistent with Zeeman splitting observations. 
(Padoan \& Nordlund 1997, 1999).\footnote{This particular result is 
supported almost exclusively by our work. A significant fraction of the 
astrophysical community still favors the traditional idea that a rather 
strong magnetic field supports MCs against their gravitational collapse. 
An example of numerical work that favors the traditional
idea of magnetic support is Ostriker, Gammie \& Stone (1999).}  
\end{itemize}  

\nocite{Scalo+98} \nocite{Vazquez-Semadeni94}

We call ``turbulent fragmentation'' the process of generation of high 
density structures by turbulent shocks. Since random super--sonic 
motions are ubiquitous in MCs, turbulent fragmentation cannot
be avoided: it is a direct consequence of the observational evidence. 
In some analytical studies it is tacitly 
assumed that turbulent fragmentation can be neglected if the kinetic 
energy of random motions is in rough equipartition with the magnetic 
energy. This assumption is wrong, because motions along magnetic field 
lines are unavoidable, and so turbulent fragmentation occurs via 
super--sonic compressions along the magnetic field lines, as discussed
by Gammie \& Ostriker (1996) and Padoan \& Nordlund (1997, 1999).

\nocite{Gammie+Ostriker96}

Numerical simulations of turbulence have been used to discuss the turbulent
origin of MC structures by Passot \& Pouquet (1987). Vazquez--Semadeni, 
Passot \& Pouquet (1996) and Ballesteros--Paredes, Hartmann \& Vazquez--Semadeni 
(1999) have used two dimensional simulations to argue that MCs are formed by 
turbulence. Padoan \& Nordlund (1997, 1999) have shown that super--sonic
and super--Alfv\'{e}nic turbulence can explain the origin of magnetized cores
in MCs, including the observed field strength--density ($B-n$) 
relation (Myers \& Goodman 1988; Fiebig \& G\"{u}sten 1989; Crutcher 1999).
The idea that proto--stellar cores and stars are formed in turbulent shocks 
has been previously discussed by Elmegreen (1993). In that work, an analysis
of the gravitational instability of the cores can be found. 
Here, we present new strong observational evidence in favor of the 
turbulent shock origin of proto--stellar cores. Such evidence
is based on the fact that dense post shock gas traces a gas component 
with a smaller velocity dispersion than lower density gas, since it maps
regions of converging flows, where the kinetic energy of the turbulent 
motion is dissipated.

\nocite{Passot+Pouquet87} \nocite{Ballesteros+99} \nocite{Myers+Goodman88} 
\nocite{Fiebig+Gusten89} \nocite{Crutcher99} \nocite{Elmegreen93}

In \S 2 and 3, numerical simulations and observational data, used in this
work, are briefly described. In \S 4 we compute the rms flow velocity
as a function of the gas density, in simulations of super--sonic turbulence,
and show that it decreases for increasing values of the gas density.
In \S 5 we show that such general property generates an observable 
relation between the rms velocity centroid and the integrated intensity 
(roughly proportional to the surface density), for the \jco\ transition,
and in \S 6 the same relation is found in the observational data. Results are 
discussed in \S 7, and conclusions are summarized in \S 8.

\section{Numerical Models}

The numerical models used in this work are based on the results
of numerical simulations of super--Alfv\'{e}nic and highly super--sonic
MHD turbulence, run on a 128$^3$ computational mesh, with periodic
boundary conditions. As in our previous works, the initial density 
and magnetic fields are uniform; the initial velocity is random, 
generated in Fourier space with power only on the large scale. 
We also apply an external random force, to drive the turbulence at a 
roughly constant rms Mach number of the flow. This force is generated 
in Fourier space, with power only on small wave numbers ($1<k<2$), 
as the initial velocity. The isothermal equation of state is used. 
Descriptions of the numerical code used to solve the MHD equations
can be found in Galsgaard \& Nordlund (1996); Nordlund, Stein 
\& Galsgaard (1996); Nordlund \& Galsgaard (1997); Padoan \& Nordlund (1999). 
 
\nocite{Nordlund+96}
\nocite{Nordlund+Galsgaard97}
\nocite{Galsgaard+Nordlund96}
\nocite{Padoan+2000AD}

In this work we neglect the effect of self--gravity, although that can be 
described with a different version of our code (Padoan et al. (2000)). 
Here we compare the relative velocity of regions of MC complexes as a 
function of their gas density or column density. Such regions are distributed 
across the full extension of the MC complexes, that is several pc. In our numerical 
models, driven continuously by an external force on the large scale, self--gravity 
is responsible for the collapse of gravitationally bound cores formed by the turbulent 
flow, but does not affect significantly the large scale flow. Since we assume 
that large scale motions in MC complexes are due to turbulence, and not to a 
gravitational collapse, we can neglect self--gravity. Similarly, we have neglected
the effect of ambipolar drift, since it is not relevant for motions on the scale
of several pc, although that is computed in a different version of our code,
using the strong coupling approximation (see Padoan, Zweibel \& Nordlund 2000). 
 
In order to scale the models to physical units, we use the following 
empirical Larson type relations, as in our previous works: \nocite{Larson81}

\begin{equation}
{\cal{M}}=2.0\left(\frac{L}{1pc}\right)^{0.5},
\end{equation} 
where $\cal{M}$ is the rms sonic Mach number of the flow, and
a temperature $T=10$~K is assumed, and
\begin{equation}
\langle n \rangle=2.0\times10^3\left(\frac{L}{1pc}\right)^{-1},
\end{equation} 
where the gas density $n$ is expressed in cm$^{-3}$.
The rms sonic Mach number is an input parameter of the
numerical simulations, and can be used to re-scale them
to physical units. When comparing theoretical models with
observations, $\cal{M}$ is in fact the only parameter that
we need to match (or its two dimensional equivalent on the maps,
$\sigma_v$ --see \S 3). In the absence of self--gravity and magnetic field,
statistical properties of turbulent flows (very large Reynolds 
number) with the same value of $\cal{M}$ should be universal, and
independent of the average density (for isothermal flows without self--gravity). 
If the magnetic field is
present, the rms Alfv\'{e}nic Mach number of the flow is also
an input parameter of the numerical simulations (it determines
the magnetic field strength). The physical unit of velocity in 
the code is the isothermal speed of sound, $C_s$, and the 
physical unit of the magnetic field is 
$C_s(4\pi\langle\rho\rangle)^{\frac{1}{2}}$ (cgs).

In this work we use two numerical models with rms velocity
3.4 and 1.7 km/s, which corresponds to ${\cal M}\approx 13.0$ 
and 6.5 respectively. Using the Larson type relations (1) and (2)
we get $L\approx 42$ and 10.5 pc and $\langle n \rangle\approx 48$ 
and 190 cm$^{-3}$. The one dimensional rms velocity for the two models
is $\sigma_v=2.0$ and 1.0 km/s, which are also recovered from the 
analysis of the synthetic spectral maps computed with these models.
The values of $\sigma_v$ have been chosen for the appropriate
comparison with the observational data presented in \S~3.\footnote{
The Taurus MC complex has $\sigma_v\approx 1.0$~km/s, and $L\approx 40$~pc,
while the Larson type relation (1) would give $L\approx 12.6$~pc.
For the purpose of this work we are interested in comparing the 
observations with numerical models with similar rms Mach number, 
and we do not try to match the physical extension of each MC complex.}
The magnetic field strength is $B\approx 5$~$\mu$G in both models.

Maps of synthetic spectra of molecular transitions are computed, using 
a non--LTE Monte Carlo radiative transfer code (Juvela 1997, 1998),
from the three dimensional density and velocity fields generated in
the numerical MHD experiments.
The method of computing synthetic spectra was presented in 
Padoan et al. (1998). For the purpose of this work we use only 
one molecular transition, namely \jco\ . Uniform temperature, $T=10$~K 
is assumed for these radiative transfer calculations, in agreement
with the isothermal equation of state used in the MHD calculations.
We are presently computing thermal equilibrium models of MCs that
we will use in a future work to study the effect of realistic 
temperature variations in molecular spectra. 

Since the spectral noise
resulting from uncertainties in our radiative transfer calculations
is always much smaller than the typical observational noise, the
comparison of synthetic spectra with observational data can be done
only after adding noise to the synthetic spectra, and the effect
of noise on the statistical properties of the spectra need to be quantified
(see \S 4).

\section{Observational Data}

We choose to use the \jco\ transition because it samples the range
of values of column density we are here interested in, and also because
several large maps of molecular clouds are available in
this transition.
We compare maps of \jco\ synthetic spectra with some of the largest
observational \jco\ spectral maps in the literature from the following 
Galactic regions: the Perseus MC complex (Billawala, Bally \& Sutherland 
1997), the Taurus MC complex (Mizuno et al. 1995), the Rosette MC complex
(Blitz \& Stark 1986; Heyer et al. 2000). These MC complexes
have an extension of approximately 30--50~pc, and radial velocity 
dispersions in the range 1--2.4~km/s. We define the total rms velocity of a
map as the rms velocity weighted with the total spectrum of the map, $T(v)$:
\begin{equation}
\sigma_v=\sqrt{\sum_v\,(v-\bar{v})^2\,T(v)\,dv\over{\sum_v T(v)\,dv}},
\end{equation}
where
\begin{equation}
\bar{v}={\sum_v v\,T(v)\,dv\over{\sum_v T(v)\,dv}}.
\end{equation}
The Blitz \& Stark map of the Rosette MC complex has $\sigma_v=2.4$~km/s,
but, limited to the region that matches the more recent Heyer et al.
map, the value is $\sigma_v=2.0$~km/s (for both maps). The full map
of the Perseus MC complex also yields a value of $\sigma_v=2.0$~km/s.
Taurus is instead much less turbulent, despite its large spatial extent,
with $\sigma_v=1.0$~km/s. 

The angular resolution is inversely proportional to the diameter of the
antenna: 4 m for the Taurus map, 7 m for the Perseus and the Blitz \& Stark
Rosette maps, and 14 m for the Heyer et al. map of Rosette.
Assuming a distance of 140 pc for Taurus, 300 pc for Perseus, and 
1600 pc for Rosette, the spatial resolution of the maps is 0.1 pc
for Taurus, 0.15 pc for Perseus, 0.84 pc for the Blitz \& Stark map
of Rosette, and 0.42 pc for the Heyer et al. map of Rosette.   
The spectral resolution is 0.1 km/s for Taurus, 0.273 km/s for Perseus,
0.68 km/s for the Blitz \& Stark map of Rosette, and 0.06 km/s for
the Heyer et al. map of the same cloud. 

Rms noise $N$ and average spectrum quality $\aq$ (see Padoan, Rosolowsky
\& Goodman 2000) also vary from 
map to map. The spectrum quality, $Q$, is related to the 
signal--to--noise, $S/N$. It is defined as the ratio of the rms 
signal (over the whole spectrum or inside a velocity window) 
to the rms noise, $N$:
\begin{equation}
Q={\sqrt{\sum_v\,T(v)^2 dv}\over{N}}
\end{equation}
The usual definition of $S/N$ is based on Gaussian fits of the spectra,
which we prefer to avoid because the \jco\ transition typically yields
spectra with significant non--Gaussian shape and multiple components. 
The spectrum quality is a sort of signal--to--noise weighted over the
whole spectrum. The relation between $Q$ and $S/N$ is discussed in
Padoan, Goodman \& Rosolowsky (2000). We call average spectrum
quality $\aq$ the value of $Q$ averaged over the whole map. Values of
$\aq$, $N$, resolutions and $\sigma_v$ are listed in Table~1 for
all the observational maps. In the following, when different maps 
are compared with each other, we make noise and velocity resolution
equal in the different maps, by adding noise and reducing the velocity 
resolution where necessary.

\section{Velocity Dispersion Versus Gas Density in Super--Sonic Turbulence}

Turbulent fragmentation generates a complex system of dense post shock sheets,
filaments and cores, reminiscent of structures observed in molecular cloud
(MC) maps. An example of a two dimensional projection of the three dimensional
density field from a simulation of super--sonic turbulence is shown in Figure~1. 
In previous works (Padoan, Jones \& Nordlund 1997; Padoan \& Nordlund 1997, 
1999), we have shown that, besides this morphological similarity, the 
density field of numerical super--sonic turbulence has important statistical 
properties in agreement with the density field of observed MCs.

Figure~2 (left panel) is a two dimensional section (no projection) of the 
same three-dimensional density field used in Figure~1. A complex system of 
filaments is apparent, with a number of high density cores inside the
filaments. In three dimensional super--sonic turbulence, filaments
are formed by two dimensional compressions (at the intersection of sheets)
and the densest cores are formed by three dimensional compressions. 
Most of the ``filaments'' in two dimensional sections like Figure~2, are 
instead two dimensional cuts through sheets, and most of the cores are local 
density maxima, due to fluctuations in the shock velocity (they
usually corresponds to strongly curved segments of filaments).
Such density maxima are often unstable to gravitational collapse,
and are the origin of proto--stellar cores. 

Since the dense gas originates in shocks, that is in regions of converging flows,
it should move with significantly lower velocity than the
lower density turbulent flow. This is illustrated in the right panel
of Figure~2, which shows the magnitude of the flow velocity on the same
plane as the left panel. Dark blue is low velocity, and dark red high
velocity. It is clear that high density filaments trace regions of low
velocity, at the intersections of high velocity ``blobs''. 
This general property of super--sonic turbulence is quantified by
the dependence of the rms flow velocity, $\sigma(v)$, on the gas density:
\begin{equation}
\sigma(v)=\langle (v-\bar{v})^2 |\rho \rangle,
\end{equation}
We repeat for different values of $\rho$, and than plot the result
versus the gas density. Expression (6) means that the
rms flow velocity is obtained as an average over the whole computational
box, using only positions where the gas density has a value contained in
an interval centered around $\rho$. The average is repeated for different
intervals of values of $\rho$, to span the whole range of densities.

The plot is shown in Figure~3, at time
$t/t_{dyn}=0.07$ (square symbols) and $t/t_{dyn}=1.0$ (asterisks),
where $t/t_{dyn}$ is the time in units of the dynamical time (defined
as the size of the computational box divided by the rms flow velocity), 
and the density is uniform in the initial conditions, at $t/t_{dyn}=0.0$.
At $t/t_{dyn}=1.0$, regions of $\rho\approx \langle \rho\rangle$ have 
$\sigma(v)$ comparable to the total rms velocity, $\sigma_v=3.0$~km/s, 
while regions with 10 times higher density have much smaller rms velocity, 
$\sigma(v)\approx 2.3$~km/sec, and at  $\rho=100\,\langle \rho\rangle$
$\sigma(v)\approx 0.9$~km/s. The rms velocity conditioned to the gas density
decreases sharply with increasing gas density already at a very early time,
$t/t_{dyn}=0.07$, when the density field is still very smooth.

\section{Rms Velocity Centroid versus Integrated Intensity}

Observational spectral maps of MCs, do not provide a direct estimate
of the three dimensional velocity field and gas density. Only radial
velocity and velocity--integrated intensity (which is roughly proportional to the
surface density), averaged along each line of sight, are directly available from the data.
However, the dependence of the rms velocity centroid on the integrated 
intensity should resemble the $\sigma(v)-\rho$ relation, since lines of
sight of high intensity are usually dominated by one or more dense cores. 
The velocity centroid, $V_0$, is the average velocity along an
individual line of sight ${\bf x}$ on the map:
\begin{equation}
V_0({\bf x})={\sum_v v\, T(v,{\bf x})dv \over {\sum_v T(v,{\bf x})dv}},
\end{equation}
where $T(v,{\bf x})$ is the signal (antenna temperature), at the velocity 
$v$ and position ${\bf x}$, and $dv$ is the width of the velocity channel.
The integrated intensity, $I({\bf x})$, is:
\begin{equation}
I({\bf x})=\sum_v T(v,{\bf x})dv.
\end{equation}

We compute the rms velocity centroid, $\sigma(V_0)$, conditioned
on $I$:
\begin{equation}
\sigma(V_0)=\langle (V_0-\bar{V_0})^2 |I \rangle_{\bf x},
\end{equation}
(spatial average) and plot it against $I$. Expression (9) means that
the rms velocity centroid is obtained as an average over the whole map,
selecting all lines--of--sight on the map with values of $I$ 
inside a given interval, [$I,I+dI$]. This rms velocity centroid
is not equivalent to a local line width, because it is obtained
as an average over the whole map. The plot is shown in 
Figure~4, for two maps of \jco\ synthetic spectra with different 
total line widths $\sigma_v$ (defined in \S 3). Figure~4 shows
that $\sigma(V_0)$ decreases significantly for increasing values of $I$.
The general property of super--sonic turbulence, namely high density 
gas moves relatively slowly, is therefore apparent also in the 
observable relation $\sigma(V_0)-I$.
The relation $\sigma(V_0)-I$ is affected by noise and by the width of
the velocity channels, which must be taken into account when comparing
different maps. Noise has been added to the synthetic 
spectra used for computing the plot in Figure~4, to yield a value of the 
spectrum quality comparable to a typical value found in the observational
data used in this work, $\aq=3.5$ (see \S 3 for the definition
of $\aq$). 

In Figure~5 we show the effect of noise (left panel) and 
spectral resolution (right panel). The effect of increasing the noise
(decreasing the value of $\aq$) is that of making the $\sigma(V_0)-I$
relation steeper, because the uncertainty in the determination of
the velocity centroids due to noise contributes to the dispersion of 
velocity centroid values, and the effect increases at decreasing values of
integrated intensity $I$. Lower spectral resolution further increases
the same effect. However, we have verified that if the noise is low
enough, $\aq>20$, both noise and velocity resolution have no effect
on the $\sigma(V_0)-I$ relation, and the squared symbols in the left 
panel of Figure~5 correspond therefore to the intrinsic relations.
We can conclude that this observable relation truly originates from 
the three dimensional $\sigma(v)-\rho$ relation, with only a partial 
contribution from noise. 

We have also verified that spatial resolution does not affect 
our results. The spatial resolution can be decreased significantly,
by rebinning the map to a smaller number of spectra, without any 
appreciable variation in the 
$\sigma(V_0)-I$ relation. As the spatial resolution is decreased,
however, the statistical sample (number of spectra) decreases, and 
statistical fluctuations (deviations around the high resolution 
$\sigma(V_0)-I$ relation) become progressively more important.

\section{Observational $\sigma(V_0)-I$ Relation}

We have shown in the previous section that the $\sigma(V_0)-I$ 
relation is sensitive to the value of the rms noise and to the
spectral resolution. In the following plots, where we compare 
different spectral maps from observations and models, we have therefore
added noise to the spectra and increased the velocity channel width $dv$
to match the map with the largest noise and $dv$. We have not modified 
the spatial resolution in any map, since that has no systematic 
effect on the $\sigma(V_0)-I$ relation, as commented above. 

The left panel of Figure~6 shows the $\sigma(V_0)-I$ relation 
for the  maps of MC complexes introduced in \S 3. For the 
Rosette MC complex we have used the portion of the full 
Blitz \& Stark map that matches the Heyer et al. map.
The two models used for the comparison have $\sigma_v=2.0$~km/s,
similar to Rosette and Perseus, and $\sigma_v=1.0$~km/s, similar 
to Taurus. It is remarkable that the Rosette and the Perseus MC 
complexes have indistinguishable $\sigma(V_0)-I$ relations,
which are also coincident with the theoretical prediction 
(square symbols), for the same value of $\sigma_v$. 
The result for Taurus is also in good agreement with the 
$\sigma_v=1.0$~km/s model. Horizontal shifts could be expected 
in the plot, since different MC complexes can have different 
surface density. However, MCs and MC complexes are known to 
approximately follow the Larson relation between density
and size (equation (2)), which implies roughly constant surface 
density, and small horizontal shifts in the plot. 

To compute the plots in the left panel of Figure~6, all maps 
have been treated to make their rms noise and velocity resolution 
equal. This is achieved by rebinning spectral profiles into a smaller 
number of velocity channels, and by adding noise, when necessary. In order
to check that the $\sigma(V_0)-I$ relations from different maps
treated in this way are really comparable, we have computed the
$\sigma(V_0)-I$ relation for the Rosette MC complex using both
the Heyer et al. map, and the portion of the Blitz \& Stark map
that matches the region covered by the Heyer et al. map. The result
is plotted in the right panel of Figure~6. The velocity resolution
of the Heyer et al. map has been decreased from $dv=0.06$~km/s 
to $dv=0.68$~km/s, and noise has been added, to match exactly 
the velocity resolution and rms noise in the Blitz \& Stark map. 
As can be seen in the right panel of Figure~6, after
this drastic treatment of the higher resolution map, the 
$\sigma(V_0)-I$ relations for the two maps are practically
indistinguishable from each other, which support the validity
of the comparison of different maps (left panel of Figure~6.
In the right panel, the case of the full Blitz \& Stark map is 
also plotted (square symbols). The full map has a higher total
rms velocity, $\sigma_v=1.4$~km/s, than the portion that matches 
the Heyer et al. map, and its $\sigma(V_0)-I$ relation is 
therefore steeper, as expected.

\section{Discussion and Conclusions}

The origin of proto--stellar cores is a fundamental problem in our 
understanding of the process of star formation. Models that have
been proposed to describe proto-stellar cores are based on i) static
or quasi--static equilibrium (e.g. Curry \& Mckee 2000; Jason \& 
Pudritz 2000), ii) thermal instability (e.g. Yoshii \& Sabano 1980; Gilden 
1984; Graziani \& Black 1987), iii) gravitational instability through
ambipolar diffusion (e.g. Basu \& Mouschovias 1994; Nakamura, Hanawa, \& 
Nakano 1995; Indebetouw \& Zweibel 2000; Ciolek \& Basu 2000), 
iv) non--linear Alfv\'{e}n waves (e.g. Carlberg \& Pudritz 1990; 
Elmegreen 1990, 1997, 1999), v) clump collisions (e.g. Gilden 1984; 
Kimura \& Tosa 1996), vi) super--sonic turbulence (e.g Elmegreen 1993; 
Klessen, Burkert and Bate 1998; Klessen, Heitsch, \& Mac Low 2000;
Padoan et al. 2000).

Many detailed comparisons between observational data and models,
which support the idea of the turbulent origin of the structure and 
kinematics of molecular clouds, have been presented in our previous 
papers (Padoan, Jones \& Nordlund 1997; Padoan et al. 1998; Padoan 
et al. 1999; Padoan \& Nordlund 1999; Padoan, Goodman \& Rosolowsky
2000). Models of numerical turbulence can be 
compared with the observations by computing i) synthetic stellar 
extinction measurements, ii) synthetic spectral maps of molecular 
transitions, iii) synthetic Zeeman splitting measurements, iv) 
synthetic polarization maps.

Maps of synthetic spectra contain a lot of information about
the kinematic, the structure and the thermal properties of 
molecular clouds, and can be analyzed with different statistical
tools. In this work we have presented a new statistical method
to analyze spectral--line maps, which is very useful because it
probes directly a very general property of super--sonic turbulence,
that is the fact that dense gas traces a gas component with a 
smaller velocity dispersion than lower density gas. This property 
arises because the gas density is enhanced in regions where the 
large scale turbulent flow converges (compressions) and the kinetic 
energy of the turbulent flow is dissipated by shocks.
If local compressions are instead due to local instabilities (e.g. 
gravitational instability, or gravitational instability mediated
by ambipolar diffusion), and the large scale motions are only the
consequence of local instabilities (as it should be in a 
self--consistent picture), gas density increases with the flow 
velocity dispersion (see for example the model by Indebetouw \& 
Zweibel 2000), contrary to the observational evidence presented 
in this work.

\nocite{Indebetouw+Zweibel2000}

The main conclusion of this work is that every model
for the origin of molecular cloud structure and proto--stellar 
cores should be tested against the $\sigma(V_0)-I$ relation.
Turbulent fragmentation provides a realistic scenario for the
origin of proto--stellar cores, which satisfies this new 
observational constraint.

\nocite{Passot+88}

\acknowledgements

We are grateful to Edith Falgarone and Phil Myers for discussions 
that stimulated this work, and to the referee for a number
of useful comments. This work was supported by NSF grant AST-9721455.
\AA ke Nordlund acknowledges partial support by the Danish National 
Research Foundation through its establishment of the Theoretical 
Astrophysics Center.

\clearpage

\onecolumn

{\bf TABLE AND FIGURE CAPTIONS:} \\

{\bf Table~1:} Cloud name; maximum spatial extension; 
total rms velocity; telescope beam size; velocity
channel width; rms noise; average spectrum quality 
(signal--to--noise); bibliographic reference. \\  
               
\begin{figure}
\centerline{\epsfxsize=13cm \epsfbox{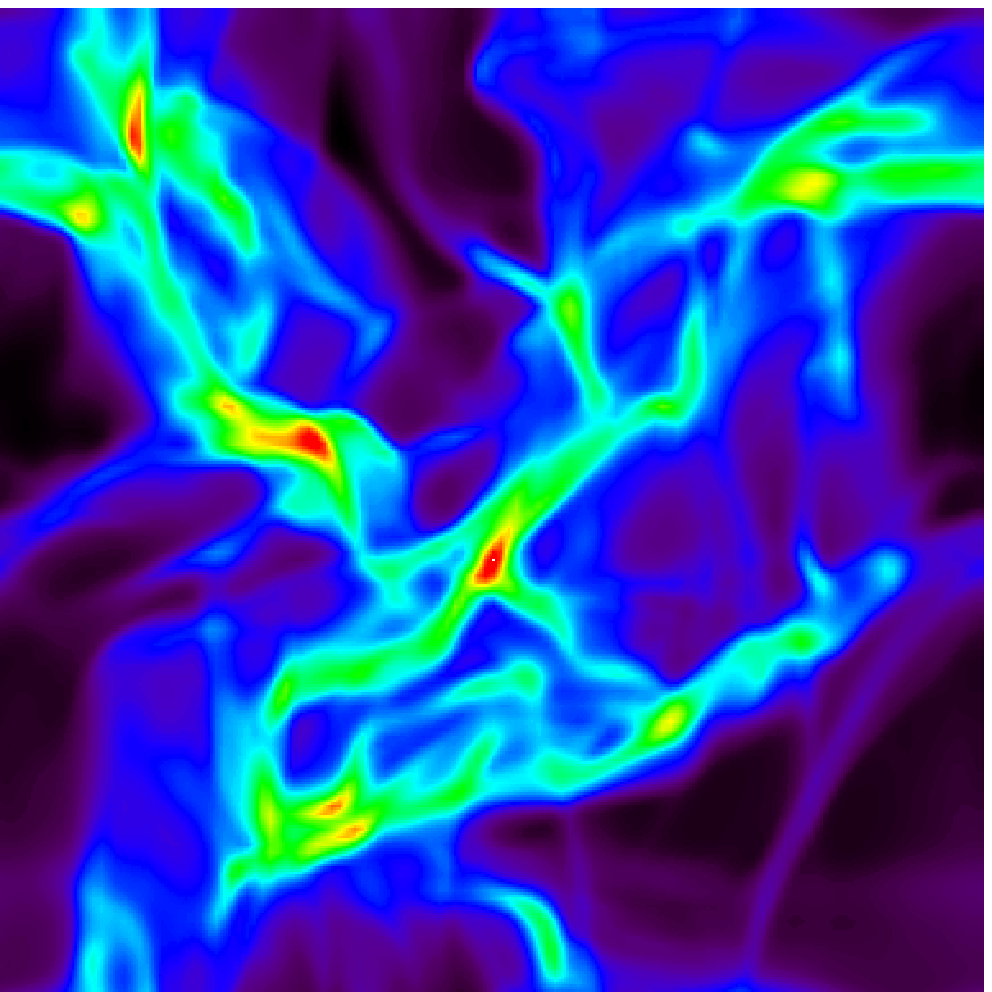}}
\caption[]{}
\label{fig1}
\end{figure}

{\bf Figure \ref{fig1}:} Two dimensional projection of the three 
dimensional density field from a simulation of isothermal
super--sonic turbulence with rms Mach number $\sim 10$. \\ 

\begin{figure}
\centerline{\epsfxsize=18cm \epsfbox{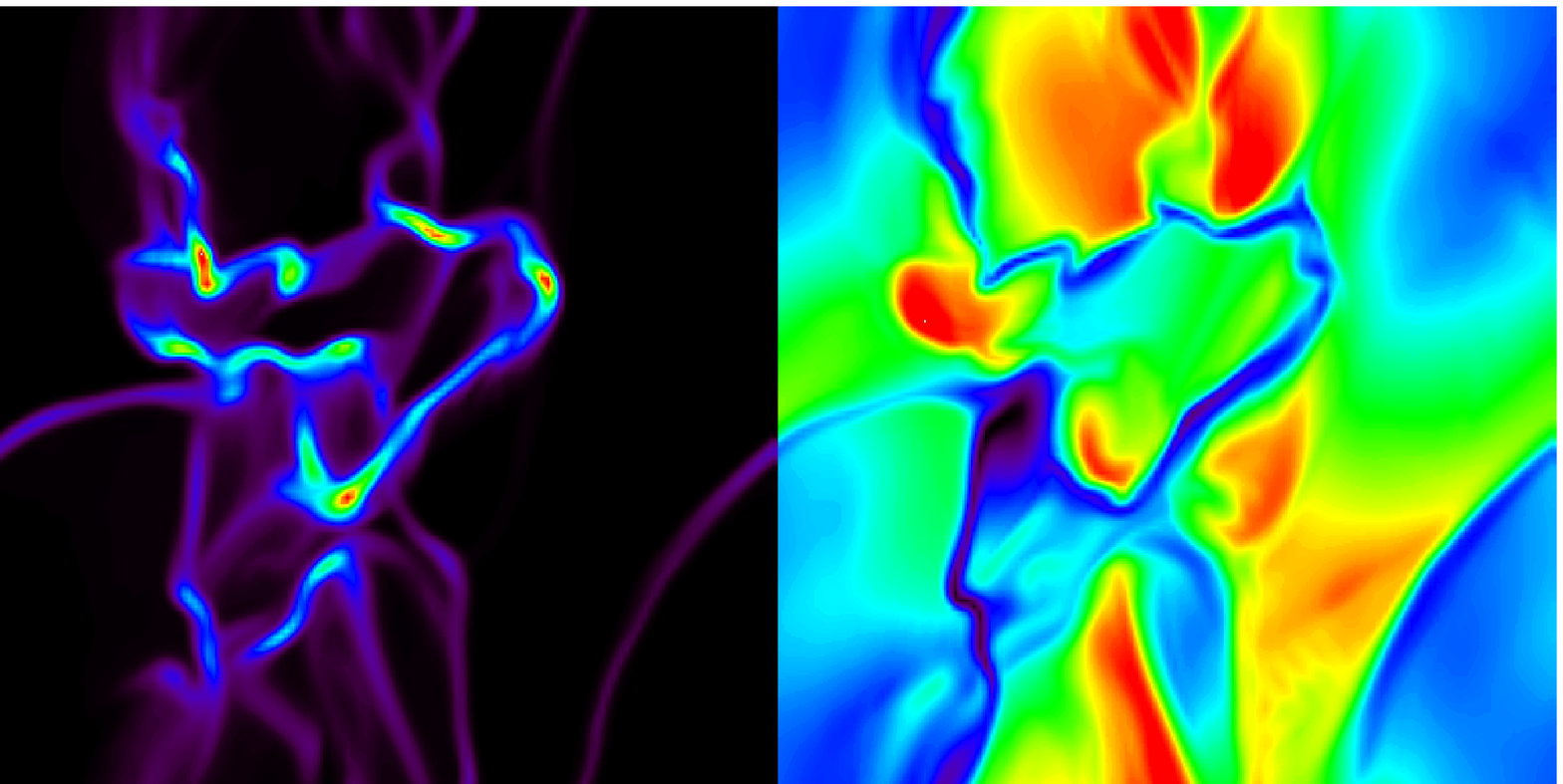}}
\caption[]{}
\label{fig2}
\end{figure}

{\bf Figure \ref{fig2}:} Left panel: Two dimensional section
(no projection) of the same density field used for Figure~1.
Most filaments are sections of sheets, and most dense cores
are density maxima inside curved segments of filaments, formed
by fluctuations in the shock velocity. Right panel: Modulus of
the three dimensional flow velocity on the same two dimensional plane as in the 
left panel. Dark blue is low velocity, and dark red high velocity.
The dense filaments on the left panel are commonly found in regions 
of small flow velocity, at the intersections of patches of high
velocity. \\

\begin{figure}
\centerline{\epsfxsize=13cm \epsfbox{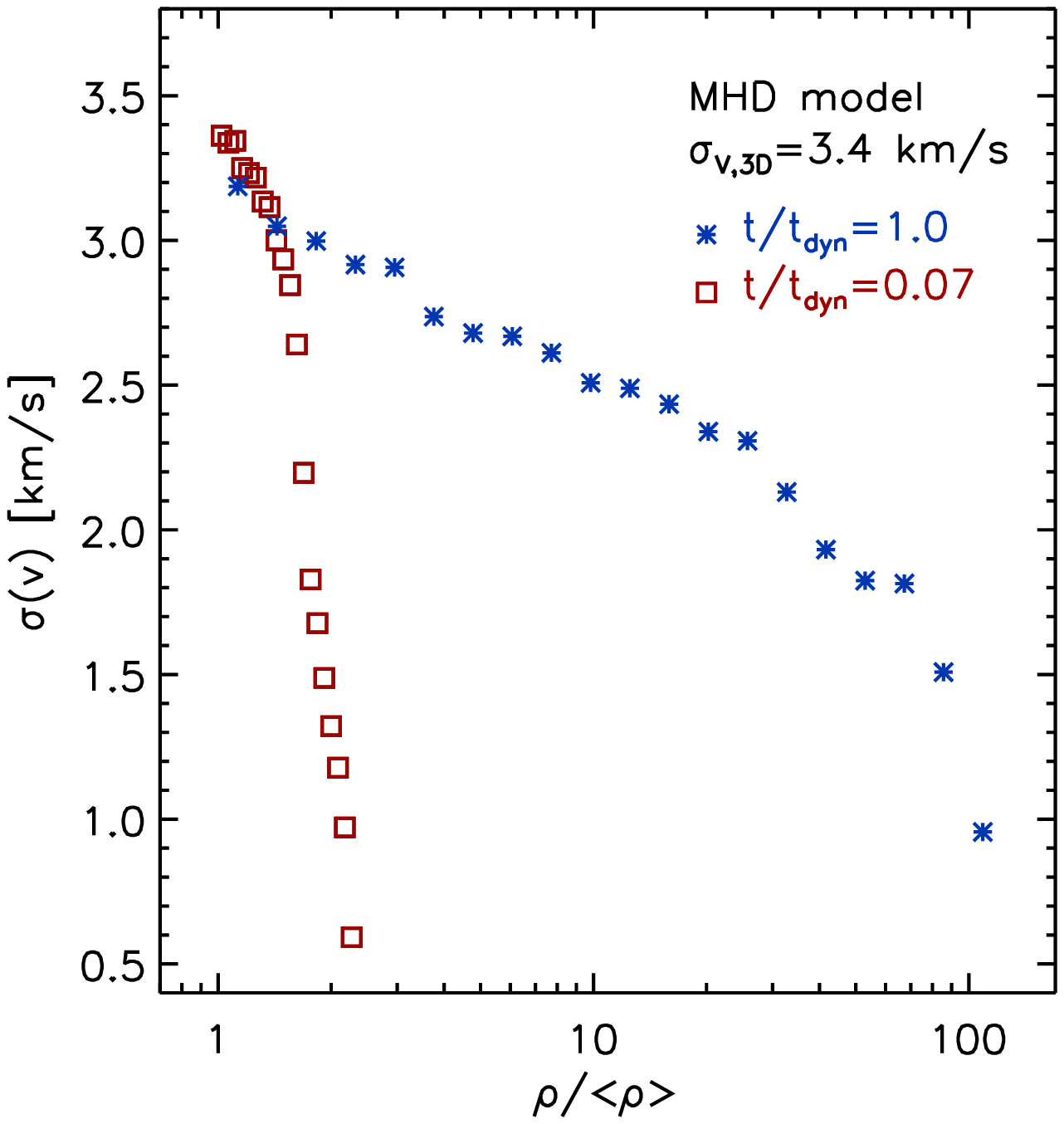}}
\caption[]{}
\label{fig3}
\end{figure}

{\bf Figure \ref{fig3}:} Rms flow velocity, conditioned to
gas density, versus the gas density, computed from a simulation
of super--sonic turbulence with rms Mach number $\sim 10$.
Squared symbols are for an early time, just 7\% of the 
dynamical time after an initial condition with uniform density;
asterisks are for a time equal to a dynamical time. Density values
are binned over 20 logarithmic intervals between the average
and the highest density. Within each interval, the density
grows by a factor of 1.04 at the early time, and 1.27 at one 
dynamical time. \\

\begin{figure}
\centerline{\epsfxsize=13cm \epsfbox{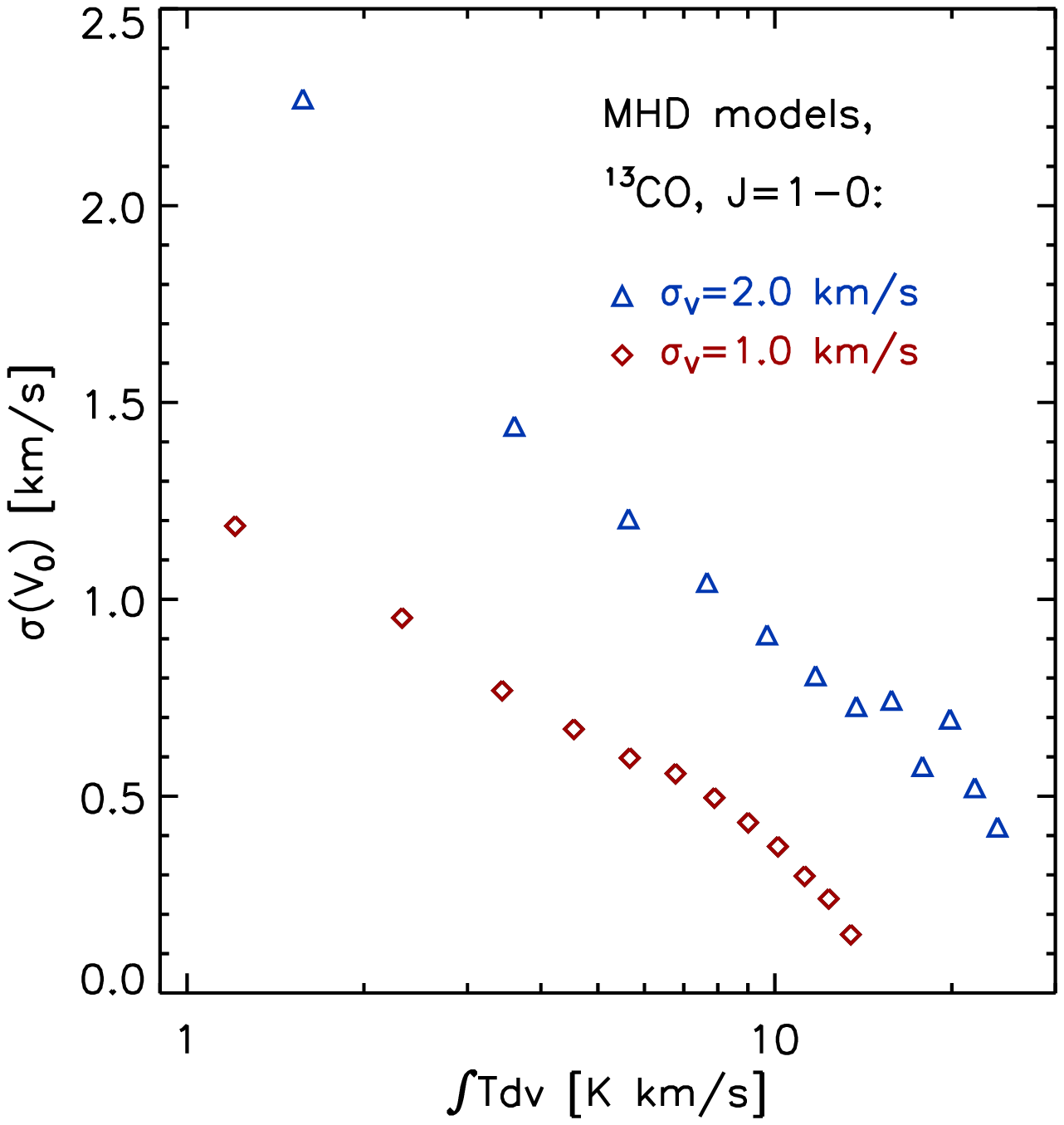}}
\caption[]{}
\label{fig4}
\end{figure}

{\bf Figure \ref{fig4}:} Rms velocity centroid, conditioned
to integrated intensity, versus the integrated intensity
(see text for details). Two maps of synthetic spectra from
MHD turbulence simulations are used, with different values
of the total rms radial velocity (intensity weighted),
$\sigma_v=1.0$~km/s (diamonds), and $\sigma_v=2.0$~km/s 
(triangles). The model with larger rms velocity has also 
a slightly larger maximum intensity (surface density), 
although both models are rescaled into physical units
using the Larson relations (see \S 2). Integrated intensity 
values are binned over 12 linear intervals between the lowest
and the highest values in the map. Within each interval, 
the density increment is $\approx 1$~K~km/s (for the 
$\sigma_v=1.0$~km/s model) and $\approx 2$~K~km/s (for the 
$\sigma_v=2.0$~km/s model). \\

\begin{figure}
{\epsfxsize=8.cm \epsfbox{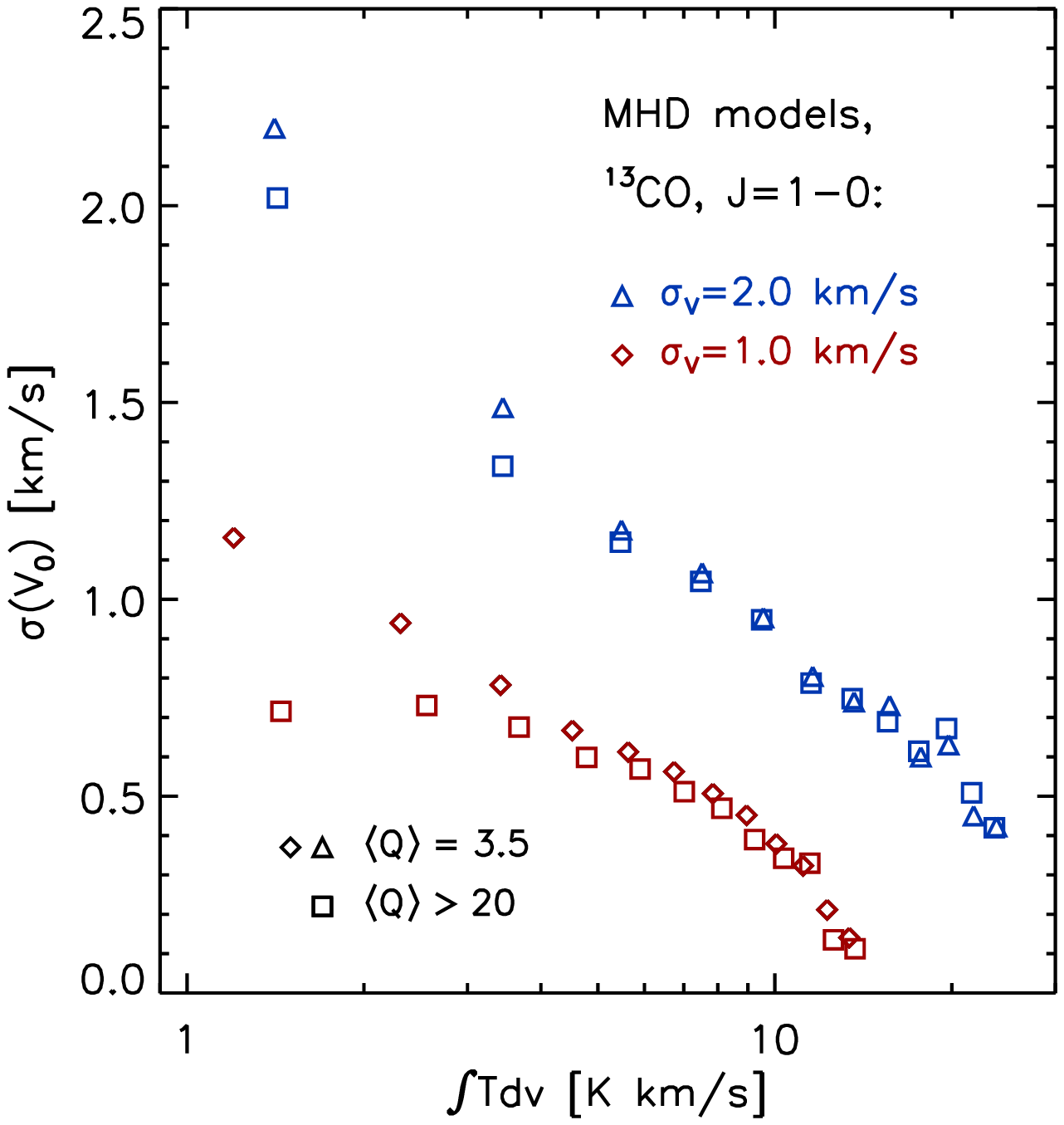}}
{\epsfxsize=8.cm \epsfbox{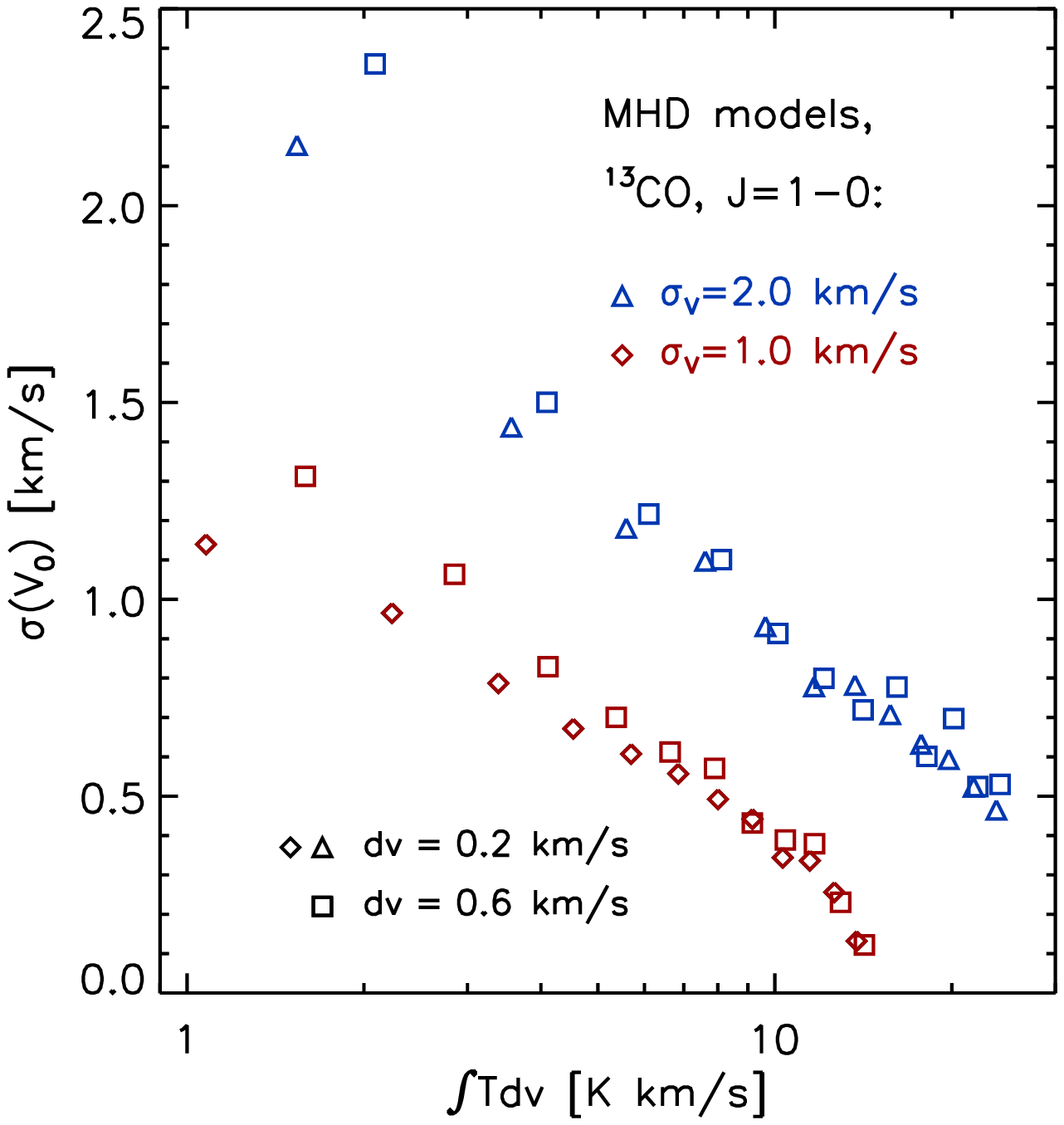}}
\caption[]{}
\label{fig5}
\end{figure}

{\bf Figure \ref{fig5}:} Effect of noise and spectral
resolution. Left panel: Conditioned rms velocity
centroid versus intensity. Triangle and diamond
symbols are the same as in Figure~4, while squared symbols
have higher spectrum quality $\aq$ (signal--to--noise). 
Different values of $\aq$ are obtained by adding different
levels of noise to the synthetic spectra. $\aq=3.5$ is typical
of the observational maps used in this work. The plot is 
not sensitive to the decreasing noise, for $\aq >20$.
Right panel: Same as left panel, but the squared symbols
are for lower spectral resolution (the velocity channels
of the synthetic spectra is increased from $dv=0.2$~km/s
to $dv=0.6$~km/s). Clearly, both the level of noise and
the spectral resolution must be taken into account when 
comparing different maps. Integrated intensity 
values are binned over 12 linear intervals between the lowest
and the highest values in the map, as in Figure~4. \\

\begin{figure}
{\epsfxsize=8cm \epsfbox{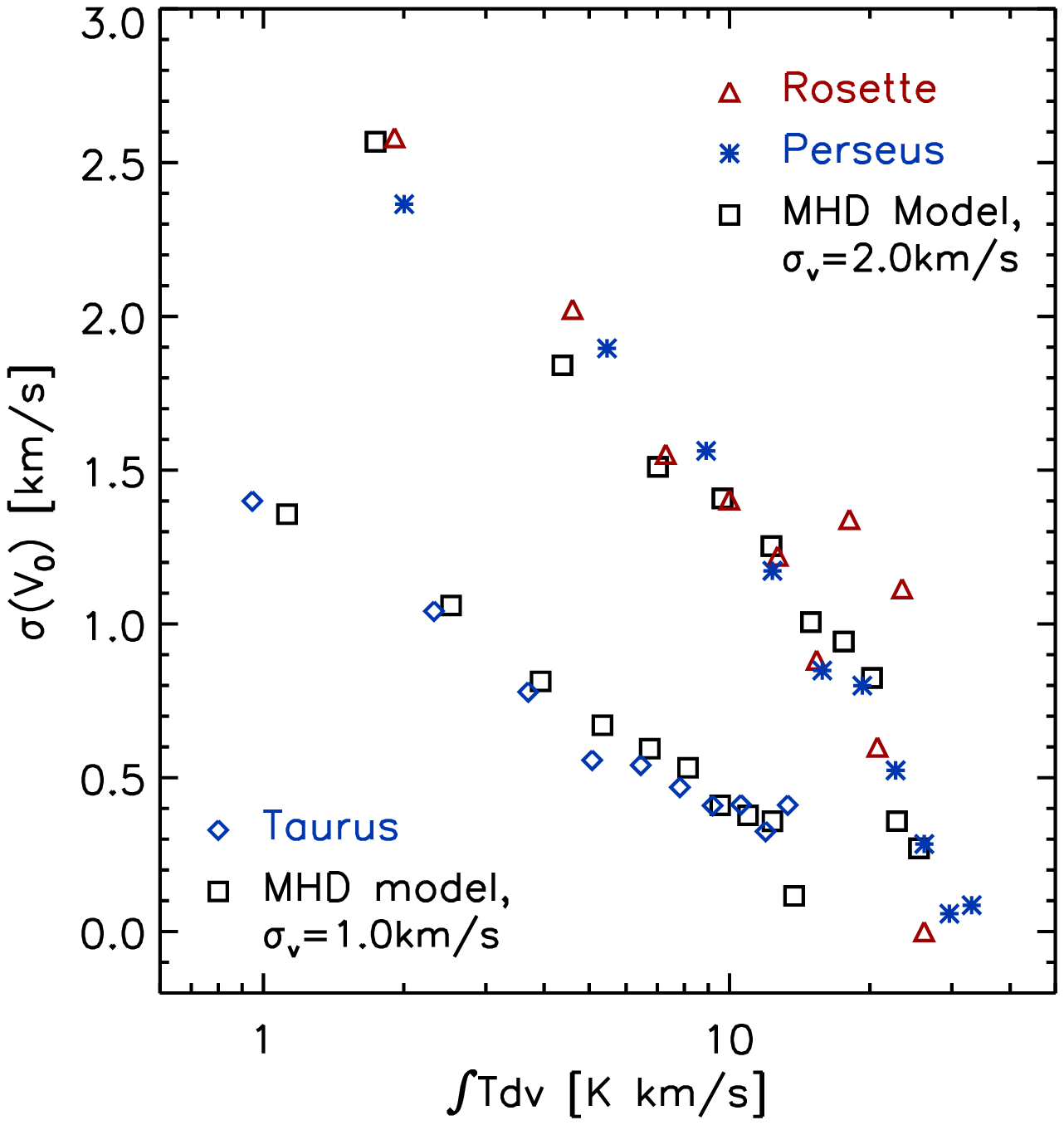}}
{\epsfxsize=8cm \epsfbox{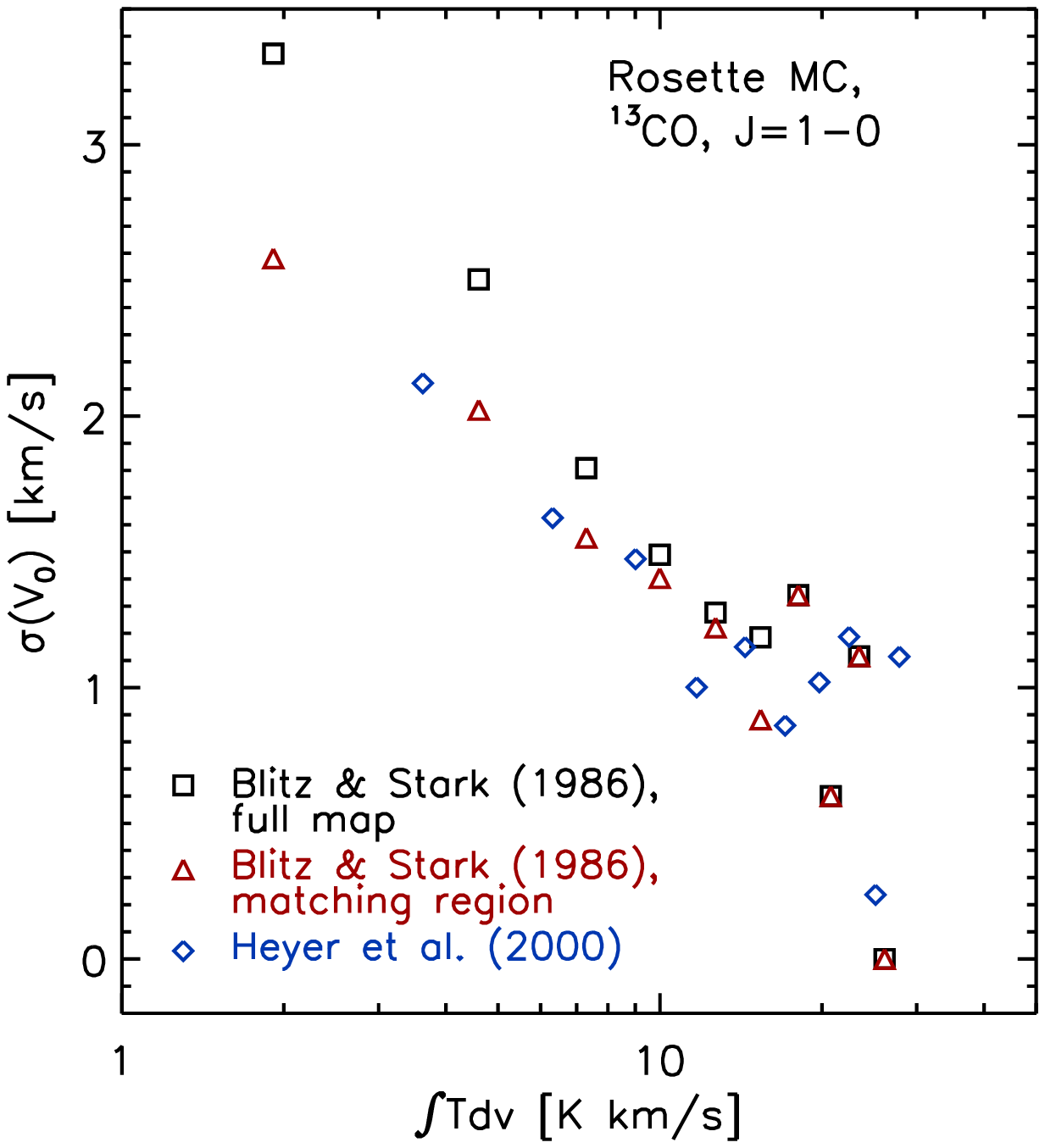}}
\caption[]{}
\label{fig6}
\end{figure}

{\bf Figure \ref{fig6}:} Observational $\sigma(V_0)-I$
relation. Left panel: Conditioned rms velocity centroid 
versus intensity for different MC complexes: Rosette, 
Perseus, and Taurus. Square symbols are for maps of 
synthetic spectra with total rms velocity $\sigma_v=1.0$~km/s
and 2.0~km/s. Right panel: Comparison of the two maps
of the Rosette MC complex (see text for details). Integrated 
intensity values are binned over 10 linear intervals between 
the lowest and the highest values in the map. \\

\clearpage
\begin{table}
\begin{tabular}{lccccccl}
\hline
\hline
MC       & $L$ [pc] & $\sigma_v [km/s]$ & $dx$ [pc]  & $dv$ [km/s] & $N$ [K] & $\aq$ & reference  \\ 
\hline
Taurus   &   38      &   1.0             &  0.10       &   0.10      &  0.24   &  2.3  & Mizuno et al. (1995)   \\ 
Perseus  &   27      &   2.0             &  0.15       &   0.27      &  0.24   &  2.8  & Billawala et al. (1997) \\ 
Rosette  &   52      &   2.4             &  0.84       &   0.68      &  0.20   &  2.1  & Blitz \& Stark (1986) \\ 
Rosette  &   36      &   2.2             &  0.84       &   0.68      &  0.19   &  4.0  & Blitz \& Stark (matching region) \\ 
Rosette  &   36      &   2.0             &  0.42       &   0.06      &  0.12   &  3.8  & Heyer et al. (2000) \\
\hline
\end{tabular}
\caption{}
\end{table}

\clearpage

\clearpage

\clearpage

\clearpage

\clearpage

\clearpage

\end{document}